\documentclass[aps,preprint,
                 amsmath,
                 amssymb,
                 showpacs,
                 superscriptaddress]{revtex4-1}
 \usepackage{etex}
\usepackage{amsmath}
\usepackage{amssymb}
\usepackage{amsfonts}
\usepackage{verbatim}
\usepackage{graphicx} 
\usepackage{color}
\usepackage{graphics}
\usepackage{pictex}
\usepackage{bm}
\usepackage{bbm}
\usepackage{pifont}
\usepackage{calc}
\usepackage{dcolumn}
\usepackage{rotating}
\usepackage{gensymb}
\usepackage{textcomp}
\usepackage{perpage}
\MakeSorted{figure}
\MakeSorted{table}

\begin{document}
\title{Dynamic Stark shift of Hydrogen-like atoms in the presence of 
       circularly polarized laser light  }
 \author{Behnam Nikoobakht}
 \affiliation{Theoretische Chemie, Physikalisch-Chemisches Institut, Universität Heidelberg, Heidelberg, Germany}
\begin{abstract}
The analytic derivation of the dynamic Stark shift of hydrogenic energy levels
in the presence of the circularly polarized laser light is presented. 
We use  the classical framework with considering an adiabatically 
damped laser+atom interaction and an approach relies on time-independent perturbation theory
with a second-quantized laser+atom dipole interaction Hamiltonian. We thus 
investigate the excitation process in atomic hydrogen or in a hydrogen like ion with 
low nuclear charge number in the presence of the circularly polarized laser light. 
The derived formula for the dynamic AC Stark shift of hydrogenic energy levels using 
these two methods are shown to be same in the limit of large photon numbers. 

\end{abstract}

\maketitle
\section{Introduction}
\label{introduction}
It is well known that the physical properties of atoms can be modified by their interaction 
with an electromagnetic radiation~\cite{Cct1972}. For instance, the virtual emission and reabsorption
of photons result in a shift of the atomic energy levels. This effect either occurs spontaneously 
(in the absence of a light source) causing to the Lamb shift, or is induced by an
external light, which result into the so-called AC Stark shift~\cite{Hbeth}. 
The AC Stark shift arises from virtual transitions induced by real photons and depends 
on the field intensity, while the Lamb shift arises from virtual transitions generated 
by virtual photons~\cite{Mbpn1994,Ak1963,Jcp2002}.

The AC Stark shift is one of the most important physical processes occurring in the 
precision spectroscopy, influencing the basic understanding of field-matter interactions, 
measurements of fundamental constants such as bound electron g factor, $1S$-$2S$ transition 
frequency~\cite{NkmH2006,Mn2000,ctm2013}. This effect is responsible for line broadening and 
frequency shift, when an atom is located in the electromagnetic field~\cite{dcho1997}.
Due to this effect, the AC Stark shift is considered the most basic systematic error associated 
in the high precision spectroscopy~\cite{MHUDJCH2006}. Due to this important rule in the high precision measurements 
of the fundamental constants, this aspect received considerable attention theoretically and 
experimentally in the literature~\cite{NkmH2006,Mn2000,ctm2013,Prsg2009,FlJc2016}.

In this investigation, our aim is to calculate the AC Stark shift due to the interaction of 
the Hydrogen-like atom with the circularly polarized laser light. We focus on the off-resonant, 
two-photon transition in which the frequency of the incident radiation is close to half of 
the atomic transition frequency. We assume that the laser light is monochromatic, implying that 
the finite band width effect of the laser light and the result dependence on the laser power are automatically 
negligible~\cite{Jcg1990A, Jcg1990B, Jcg1990C}. This paper is organized as follow. In Sec.~\ref{classic}, 
the dynamic Stark shift for Hydrogen-like atom in the presence of the circularly polarized laser light 
is dealt with in the framework of time-dependent perturbation theory. In Sec.~\ref{secondquantization}, 
we applied the semiclassical description, in which Hydrogen-like atom is described in the framework 
of the quantum mechanics and the laser field is characterized in the framework of the classical mechanics 
and showed that a fully quantized-field treatment goes to a classical result when the photon number in 
the laser mode is large enough. Finally, the summary 
and concluding remarks were presented in Sec.~\ref{summ}.

\section{Classical approach}
\label{classic}
Consider a Hydrogen-like atom (ion) that is adiabatically embedded in the remote past
and future in circularly polarized laser field

\begin{align}
{\bm E}(t)=\frac{\varepsilon_L e^{-\epsilon \vert t \vert }}{\sqrt{2}}\left[{\bm e}_x \cos(\omega_L t)+{\bm e}_y \sin(\omega_L t)\right],
\label{filed}
\end{align}
where the ${\bm e}_x$ and ${\bm e}_y$ refer to the unit vectors and $\omega_L$ is the laser frequency.
We have assumed that the field energy density is normalized as $\varepsilon_L^2$.
$\epsilon$ is an infinitesimal damping parameter. Here, we have adopted the dipole 
approximation which implies that the wavelength of the radiation is far greater than 
the atom size. The evolution of the unperturbed Hamiltonian of the atom $H_0$ with the 
eigenfunction $\vert \phi \rangle$ in the monochromatic laser field is determined by
\begin{align}
i\hbar \frac{\partial \psi}{\partial t}= H \psi,
\label{shrod} 
\end{align}
where the Hamitonian of the atom $+$ laser reads
\begin{align}
 H&=H_0+V(x,y,t)
\label{hamilto}
\end{align}
In Eq.~(\ref{hamilto}), $H_0$ is 
\begin{align}
H_0&= \frac{{\bm p}^2}{2m_e}-\frac{Ze^2}{4\pi\varepsilon_0 r},
\label{hydroHami} 
\end{align}
where Z refers t othe nuclear charge and $V (x, y, t)$ describes the interaction of the Hydrogen-like 
atom in the presence of the circularly polarized laser light (in the length gauge),
\begin{align}
V (x, y, t)&= e^{-\epsilon \vert t \vert }\left[V(x)\cos(\omega_L t)+V(y)\sin(\omega_L t)\right]
\label{poten}
\end{align}
with $V(x)=-e\varepsilon x/\sqrt{2}$ and $V(y)=-e\varepsilon y/\sqrt{2}$. In this investigation, 
we shall take into account only the solutions of Eq.~(\ref{shrod}) corresponding the adiabatic 
switching of the perturbation $V(x,y,t)$ at the remote past and future ($t \rightarrow \pm \infty$). 
This indicates that the solution of Eq.~(\ref{shrod}) at $t \rightarrow \pm \infty$ turns into the 
eigenstate of the unperturbed Hamiltonian $H_0$, which for simplicity shall be assumed to be 
nondegenerate.

To describe the effect of the off-resonance perturbation by a time-dependent electric
field Eq.~(\ref{filed}) on the reference state $\vert \phi \rangle$ of the bare atom, 
we use the interaction picture, in which the field-atom interaction is represented as
\begin{align}
V_I&=e^{\frac{i}{\hbar}H_0 t} V(x,y,t) e^{-\frac{i}{\hbar}H_0 t},
\label{poteInte}
\end{align}
From which we obtain the expansion of the time evolution operator in the interaction
picture
\begin{align}
U_I(x,y,t)&=T \exp\left[-\frac{i}{\hbar} \int_{-\infty}^t U_I(\tau) d\tau\right]\nonumber\\
 &=1-\frac{i}{\hbar} \int_{-\infty}^t dt'V_I(x,y,t)+
\left(\frac{-i}{\hbar}\right)^2\int_{-\infty}^t dt'\int_{-\infty}^{t'} dt''V_I(x,y,t') V_I(x,y,t'')
\label{operator} 
\end{align}
where $T$ is the time-ordering operator. Due to the laser-atom interaction, the reference
state $\vert \phi \rangle$ evolves into a time dependent atomic state $\vert \psi_I(t)\rangle$,
\begin{align}
\vert \psi_I(x,y,t) \rangle&=U(x,y,t)\vert \psi (x,y,t=-\infty)\rangle=U(x,y,t)\vert \phi \rangle.
\label{operator1}
\end{align}
Since the interaction is weak compared to the Coulomb field of atom, we may expand $\vert \psi_I \rangle$ 
in a complete set $\{\vert m_i\rangle\}$ of eigenstates of $H_0$,
\begin{align}
\vert \psi_I\rangle&= \sum_m c_m(t) \vert m \rangle,
\label{wavwfun} 
\end{align}
where $c_m(t)=\langle m\vert \psi_I(t)\rangle$ are expansion coefficients, that satisfy the initial 
condition $c_\phi(-\infty)=0$ and $c_m(-\infty)=0 ~(m\neq \phi)$. To obtain the energy shift and 
ionization rate of the reference state, we calculate the projection 
\begin{align}
 c_\phi(t)=\langle \phi \vert \psi_I(t) \rangle =\langle \phi \vert U_I(r,t) \vert \phi \rangle.
 \label{coffe}
\end{align}
We now substitute Eqs. (\ref{poten}), (\ref{poteInte}) and (\ref{operator1}) into Eq. (\ref{coffe}). 
Since $\langle \phi \vert x\vert \phi \rangle=\langle \phi \vert y\vert \phi \rangle=0$, the expansion 
may be taken up to the second order,
\begin{align}
c_\phi(t)&1-\frac{1}{\hbar^2}M, 
\label{coffe1}
\end{align}
where
 \begin{align}
M&=\int_{-\infty}^t dt' \int_{-\infty}^{t'} \langle \phi V_I(x,y,t')V_I(x,y,t'')\vert \phi \rangle\nonumber\\
&=\sum_m\int_{-\infty}^t dt'\int_{-\infty}^{t'}dt'' \langle \phi \vert V_I(x,y,t')\vert m\rangle 
\langle m \vert V_I(x,y,t'')\vert \phi\rangle
\label{coffe2} 
\end{align}
The sum counts all bound and continuum states. Since the time dependence of the potentials
is harmonic, the integrations can be easily worked out and thus Eq.(\ref{coffe2}) reads
\begin{align}
M&=-\frac{\hbar}{4i}\sum_{m,\pm}\frac{e^{2\epsilon t}}{2\epsilon}
\bigg[\frac{\langle \phi \vert V(x) \vert m \rangle \langle m \vert V(x) \vert \phi \rangle}{E_\phi-E_m\pm \hbar\omega_L
+i\hbar\epsilon}+\frac{\langle \phi \vert V(y) \vert m \rangle \langle m \vert V(y) \vert \phi \rangle}{E_\phi-E_m\pm \hbar\omega_L+i\hbar\epsilon}\nonumber\\
&\pm i\frac{\langle \phi \vert V(x) \vert m \rangle \langle m \vert V(y) \vert \phi \rangle}{E_\phi-E_m\pm \hbar\omega_L
+i\hbar\epsilon}\mp i\frac{\langle \phi \vert V(y) \vert m \rangle \langle m \vert V(x) \vert \phi \rangle}{E_\phi-E_m\pm \hbar\omega_L+i\hbar\epsilon}\bigg]
\label{coffe3}
\end{align}
Inserting this result into Eq. (\ref{coffe1}), and considering
\begin{align}
 \frac{\partial}{\partial t} \ln c_\phi(t)&=-\frac{i}{4\hbar}\sum_{m,\pm}\frac{e^{2\epsilon t}}{2\epsilon}
\bigg[\frac{\langle \phi \vert V(x) \vert m \rangle \langle m \vert V(x) \vert \phi \rangle}{E_\phi-E_m\pm \hbar\omega_L
+i\hbar\epsilon}+\frac{\langle \phi \vert V(y) \vert m \rangle \langle m \vert V(y) \vert \phi \rangle}{E_\phi-E_m\pm \hbar\omega_L+i\hbar\epsilon}\nonumber\\
&\pm i\frac{\langle \phi \vert V(x) \vert m \rangle \langle m \vert V(y) \vert \phi \rangle}{E_\phi-E_m\pm \hbar\omega_L
+i\hbar\epsilon}\mp i\frac{\langle \phi \vert V(y) \vert m \rangle \langle m \vert V(x) \vert \phi \rangle}{E_\phi-E_m\pm \hbar\omega_L+i\hbar\epsilon}\bigg],
\label{coffe4}
\end{align}
here $e^{2\varepsilon t}$ has been replaced by 1. The solution of Eq.~(\ref{coffe4}) is
\begin{align}
c_\phi(t)&= e^{-\frac{i}{\hbar}\Delta E_{AC}(\phi)t},
\label{coffe5}
\end{align}
where we have defined the dynamic stark shift $\Delta E_{AC}(\phi)$ of the reference state $\vert \phi \rangle$,
\begin{align}
 \Delta E_{AC}(\phi)&=\frac{1}{4}\sum_{m,\pm}\bigg[\frac{\langle \phi \vert V(x) \vert m \rangle \langle m \vert V(x) \vert \phi \rangle}{E_\phi-E_m\pm \hbar\omega_L
 +i\hbar\epsilon}+\frac{\langle \phi \vert V(y) \vert m \rangle \langle m \vert V(y) \vert \phi \rangle}{E_\phi-E_m\pm \hbar\omega_L
 +i\hbar\epsilon}\nonumber\\
&\pm i\frac{\langle \phi \vert V(x) \vert m \rangle \langle m \vert V(y) \vert \phi \rangle}{E_\phi-E_m\pm \hbar\omega_L
+i\hbar\epsilon}\mp i\frac{\langle \phi \vert V(y) \vert m \rangle \langle m \vert V(x) \vert \phi \rangle}{E_\phi-E_m\pm \hbar\omega_L+i\hbar\epsilon}\bigg].
\label{coffe6}
\end{align}
Considering Eq.~(\ref{wavwfun}), the wavefunction of the system in the Schr\"{o}dinger picture reads
\begin{align}
 \vert \psi(t)&=e^{-\frac{i}{\hbar}H_0 t}\vert \psi_I(t) \rangle=e^{-\frac{i}{\hbar}H_0 t} \left[c_\phi(t)\vert \phi \rangle+
\sum_{m\neq \phi} c_m(t)\vert m \rangle \right ].
\label{coffe7}
\end{align}
Thus we obtain the projection
\begin{align}
\langle \phi \vert \psi (t) \rangle =e^{-\frac{i}{\hbar}[E_\phi+\Delta E_{AC}]t},
\label{coffe8}
\end{align}
which shows the perturbative effect on the time-evolution of the reference state. 
$\Delta E_{AC}(\phi) $ is generally complex arising from the fact that the atomic 
system may exchange virtual and real photons with the laser field in the course 
of interaction (see below, Eq.~(\ref{coffe21})).
Introducing the complex coordinates
\begin{align}
 x_{+1}&=-\frac{1}{\sqrt{2}}(x+iy)\nonumber\\
 x_{-1}&=\frac{1}{\sqrt{2}}(x-iy)
\label{coffe10}
\end{align}
we may cast the dynamic Stark shift of Eq.~(\ref{coffe6}) into a more concise form
\begin{align}
\Delta E_{AC}(\phi)=-\frac{e^2\varepsilon_L^2}{4}\sum_\pm \langle \phi \vert x_{\mp 1}\frac{1}{H_0-E_\phi \pm \hbar \omega_L}x_{\pm 1} \vert \phi \rangle
\label{coffe11}
\end{align}
In this relation, we have used the closure relation over the spectrum of $H_0$. If we define the dynamic
polarizability
\begin{align}
 P_{\omega_L}(\phi)=\sum_\pm \langle \phi \vert x_{\mp 1}\frac{1}{H_0-E_\phi \pm \hbar \omega_L}x_\pm \vert \phi \rangle,
\label{coffe12}
\end{align}
then the dynamic Stark shift reads,
\begin{align}
\Delta E_{AC}(\phi)=-\frac{1}{4}e^2\varepsilon_L^2 P_{\omega_L}(\phi)=-\frac{e^2}{2\varepsilon_0 c}I_L P_{\omega_L}(\phi),
\label{coffe12a}
\end{align}
where $I_L = \varepsilon_0 c \varepsilon_L^2/2$ is the intensity of the laser field. 
Eq.~(\ref{coffe12}) can be interpreted as the sum of the amplitudes of two-photon processes. 
In the first process the atom absorbs one quantum and goes into a virtual state (the term 
involving $-$ sign in the denominator) and after emitting the same $\hbar \omega_L$ quantum 
it returns to the initial state. In the second process (the term involving $+$ sign in the 
denominator) the atom emits the $\hbar \omega_L$ quantum first and goes into the virtual 
state and after absorbing the same quantum, it returns to the initial state. The result is 
the shift of an energy level.

We would like to mention that in the evaluation of
the matrix element of Eq.~(\ref{coffe12}), the contribution of two intermediate states 
with energies $E_\phi-\hbar \omega_L$ and $E_\phi+\hbar \omega_L$ must be summed. 
For the evaluation of these energies, one needs to set the frequency of the laser 
field $\omega_L=(E_e-E_g)/2\hbar$ in the two-photon resonant spectroscopy of 
the transition $g \Leftrightarrow e$. Since in the study of dynamic Stark shift 
the final state is in the continuum, energies of the intermediate states are 
always positive. Thus, the dynamic polarizability of Eq.~(\ref{coffe12}) gains 
an imaginary part. It characterizes the population loss rate (decay rate) due to 
ionization, while the real part of the dynamic polarizability is associated with the 
level shift $\vert \phi \rangle$,
\begin{align}
\text{Re}\left [\Delta E_{AC}(\phi)\right ]&= \hbar I_L \beta_{\text{AC}}(\phi)\nonumber\\
\text{Im}\left [\Delta E_{AC}(\phi)\right ]&=-\pi \hbar I_L \beta_\text{ioni}(\phi),
\label{coffe21}
\end{align}
where $\beta_\text{AC}$ and $\beta_\text{ioni}$ are called dynamic Stark shift and ionization
coefficients, respectively. The decay rate $\gamma_i(\phi)$ and ionization cross section 
$\sigma_i(\phi)$ as a function the ionization coefficient
reads 
\begin{align}
 \gamma_i(\phi)&=2\pi \beta_\text{ioni}(\phi) I_L\nonumber\\
 \sigma_i(\phi)&= 2\pi \hbar \omega_L\beta_\text{ioni}(\phi).
\label{coffe22}
\end{align}
The dynamic Stark shift coefficients and total correction $\sigma_{\text{ioni}}$ in Eq.~(\ref{coffe22}) 
can be scaled by $\frac{1}{Z^4}$ and $\frac{1}{Z^2}$, respectively and thus one can generalize the 
results to any Hydrogen-like atoms (ions) up to $Z=11$.
\section{Treatment in the second quantization}
\label{secondquantization}
In the preceding section, we dealt with the interaction between Hydrogen-like atomic energy
levels and the electromagnetic radiation and assumed that the field to be classical. In this
section, in view of light being a photon field, we discuss the interaction of the quantized
radiation field with a Hydrogen-like atom. 
\subsection{Atom and quantized radiation interaction Hamiltonian}
The interaction of a monochromatic laser mode of angular frequency $\omega_L$ with a 
Hydrogen-like atom can be formulated in a general way [see {\it e.g.}, Ref.~\cite{Mos1997,Mfox2006}]. 
One can characterize the Hamiltonian corresponding to the atom $+$ laser system in terms 
of three parts as follows,
\begin{align}
H_I=H_0+\hbar \omega_L+\hbar\omega_L a_L^\dag a_L+H_L,
\label{coffe13}
\end{align}
where the first two terms are the Hydrogen-like atom $H_0$ and the quantized field Hamiltonians,
respectively. In Eq.~(\ref{coffe13}), $H_0$ reads 
\begin{align}
 H_0=\sum_m\vert m \rangle \langle m \vert +\frac{{\bm p}^2}{2m_e}-\frac{Ze^2}{4\pi \varepsilon_0 r}
\label{coffe14}
\end{align}
where $\vert m \rangle$ represents a complete set of atomic energy eigenstates including both 
the discrete and continuum region of the spectrum ($m_e$ and $e$ are electron mass and charge,
respectively). Since we restrict ourselves to a single-mode field with the frequency $\omega_L$, 
the corresponding Hamiltonian of the laser field can be written in terms of the creation and
annihilation operators $a_L^\dag$ and $a_L$, respectively. The corresponding eigenstates are called
Fock states [21]. The energy eigenvalues are discrete, in contrast to the classical
electromagnetic theory, where the energy is continuous. The last part of the Eq.~(\ref{coffe13}) 
denotes the perturbed term of the Hamiltonian which is responsible for the interaction between a 
Hydrogen-like atom and the monomode electromagnetic radiation. In the framework of the length gauge 
and dipole approximation, this term $H_L$ reads
\begin{align}
H_L&=-e {\bm r}\cdot {\bm E}
\label{coffe15}
\end{align}
In view of the fact that the laser field is polarized circularly, thus
$H_L$ reads
\begin{align}
 H_L=-e \varepsilon_L \left [-x_{+1}a_L+x_{-1}a_{L}^\dag \right],
\label{coffe16}
\end{align}
where $\varepsilon_L=\sqrt{\hbar \omega_L/2\epsilon_0 V}$, $V$ is the normalization volume 
and $x_{+1}$ and $x_{-1}$ were introduced in Eq.~(\ref{coffe10}). Eq.~(\ref{coffe10}) 
implies that $x_{+1}^\dag = -x_{-1}$ and $x_{-1}^\dag = -x_{+1}$ indicating that the 
Hamiltonian $H_L$ is hermitian. 

\subsection{Quantized field approach}
\label{quantizationfieldapproach}
We are now in a position to discuss the interaction between circularly polarized laser light
and the Hydrogen-like atom from the quantum nature point of view of the radiation field
using Eqs.~(\ref{coffe13}) and (\ref{coffe16}). Working in the Schr\"{o}dinger picture, 
perturbation theory leads to the second-order energy shift of the unperturbed eigenstate 
$\vert \phi, n_L \rangle$,
\begin{align}
\Delta E_{AC}(\phi)&= \sum_m \bigg[ \frac{\langle \phi, n_L \vert H_L \vert m, n_L-1\rangle 
\langle m, n_L-1 \vert H_L \vert \phi, n_L\rangle}{E_\phi+n_L\hbar\omega_L-(E_m+(n_L-1)\hbar\omega_L)}\nonumber\\
&+\frac{\langle \phi, n_L \vert H_L \vert m, n_L+1\rangle 
\langle m, n_L+1 \vert H_L \vert \phi, n_L\rangle}{E_\phi+n_L\hbar\omega_L-(E_m+(n_L+1)\hbar\omega_L)}\bigg].
\label{coffe17}
\end{align}
Inserting Eq.~(\ref{coffe16}) into Eq.~(\ref{coffe16}), $\Delta E_{AC}(\phi)$ reads
\begin{align}
 \Delta E_{AC}(\phi)=\frac{e^2\hbar \omega_L}{2\varepsilon_0 V}\sum_m
\bigg[\frac{\langle \phi \vert x_{-1} \vert m \rangle \langle m \vert x_{+1} \vert \phi \rangle}{E_\phi-E_m+\hbar \omega_L}n_L+
\frac{\langle \phi \vert x_{+1} \vert m \rangle \langle m \vert x_{-1} \vert \phi \rangle}{E_\phi-E_m-\hbar \omega_L}(n_L+1)\bigg].
\label{coffe18}
\end{align}
Eq.~(\ref{coffe18}) derived using the time-dependent field operator is suitable  for the represention 
the dynamic AC stark shift of the coupled system, atom+laser field when the photon number $n_L$ is not large.
In nonrelativistic quantum mechanics, the classical description is trustworthy whenever
the noncommutativity of dynamical variables is unimportant [see, {\it e.g.}, textbook ~\cite{Emer1970}].
Likewise, in the quantum theory of radiation, if we could ignore the right-hand-side of
$[a_L, a_L^\dag] = 1$ then we would return to the classical description. Due to the proportionality
the nonvanishing matrix elements of $a_L$ and $a_L^\dag$ to the occupation number $\sqrt{n_L}$, 
the classical description Eq.~(\ref{coffe12}) can be obtained again from Eq.~(\ref{coffe18}) in 
the limit of a large occupation number. In other words, in the classical limit, $n_L\rightarrow \infty$ 
,$V \rightarrow \infty$, whereas $n_L/V =\text{const.}$, Eq.~(\ref{coffe18}) becomes
\begin{align}
\Delta E_{AC}(\phi)=-\frac{e^2n_L\hbar\omega_L}{2\varepsilon_0 V} P_{\omega_L}(\phi),
\label{coffe19}
\end{align}
where $P_{\omega_L}(\phi)$ is given by Eq.~(\ref{coffe12}). Comparing Eq.~(\ref{coffe19}) with 
the classical result in Eq.~(\ref{coffe12}), we find 
\begin{align}
 I_L= \varpi c,
\label{coffe20}
\end{align}
where $\varpi=n_L\hbar \omega_L/V$ is the energy density of the field in the quantum version.

\section{Summary and conclusion}
\label{summ}
The analytic relation for dynamic stark shift of hydrogen-like atom was obtained in two approaches. 
In the first treatment (the classical method), it was assumed that the hydrogen-like atom is shined by 
the circularly laser light, where Hydrogen-like atom adiabatically embedded in the remote past
and future in this field. In the second treatment, the dynamic stark shift was studied in the view of 
the fact that the radiation field being quantized. The later approach concurs with the classical field 
description in the limit of high photon density. This (quantized field) approach showed that the AC stark shift results
from self-energy like formalism when the sum over virtual modes of the photon field is limited to a 
single mode thus the AC stark shift can be  expressed as a stimulated radiative correction.



\begin{thebibliography}{unsrt}


\bibitem{Cct1972}
C. Cohen-Tannoudji, R. Dupont-Roc, and G. Grynberg Phys. Rev. A \textbf{5} 968 (1972).

\bibitem{Hbeth}
H. A. Bethe and E. E Salpeter {\it Quantum Mechanics of One and Two Electron Atioms} 
(Springer, Berlin and Academic press, New York, $1957$).

\bibitem{Mbpn1994}
M. Brune, P. Nussenzveig, F. Schmidt-Kaler, F. Bernardot, A. Maali, J. M. Raimond, and 
S. Haroche Phys. Rev. Lett. \textbf{72}, 3339 (1994).

\bibitem{Ak1963}
A. Kastler J. Opt. Soc. Am. \textbf{53}, 902 (1963).

\bibitem{Jcp2002}
J. Camparo and P. Lambropoulos J. Opt. Soc. Am. B \textbf{19}, 1169 (2002).

\bibitem{Sakurai}
J. J. Sakurai, {\it Modern Quantum Mechanics} (Addison-Wesley Publishing Company, Inc. $1985$). 

\bibitem{NkmH2006}
N. Kolachevsky, M. Haas, U. D. Jentschura, M. Herrmann, P. Fendel, M. Fischer, R. Holzwarth, Th. Udem, C. H. Keitel, and T. W. H\"{a}nsch, 
Phys. Rev. A \textbf{74}, 052504 (2006).

\bibitem{Mn2000}
M. Niering {\it et al.}, Phys. Rev. Lett. \textbf{84}, 5496 (2000).

\bibitem{ctm2013}
C. T. M. Weber, N. Leefer, and D. Budker, Phys. Rev. A \textbf{88}, 062503 (2013).

\bibitem{dcho1997}
D. Cho {\it et al} Phys. Rev. A  \textbf{55}, 1007 (1997).

\bibitem{MHUDJCH2006}
M.~Haas, U.~D.~Jentschura, C~.H.~Keitel,  Am. J. Phys. \textbf{74}, 77 (2006).

\bibitem{Prsg2009}
P. Rosenbusch, S. Ghezali, V. A. Dzuba, V. V. Flambaum, K. Beloy, and A. Derevianko, Phys. Rev. A \textbf{79}, 013404 (2009).

\bibitem{FlJc2016}
F. Levi, J. Camparo, B. Francois, C. E. Calosso, S. Micalizio, and A. Godone
Phys. Rev. A \textbf{93}, 023433 (2016).


\bibitem{Jcg1990A}
J. C. Garreau, M. Allegrini, L. Julien, and F. Biraben J. Phys. (France) \textbf{51} (1990) 2263.

\bibitem{Jcg1990B}
J. C. Garreau, M. Allegrini, L. Julien, and F. Biraben J. Phys. (France) \textbf{51} (1990) 2275.

\bibitem{Jcg1990C}
J. C. Garreau, M. Allegrini, L. Julien, and F. Biraben J. Phys. (France) \textbf{51} (1990) 2293.

\bibitem{Mos1997}
M. O. Scully and M. S. Zubairy, Quantum optics . ( Cambridge university, Cambridge, U.K., 1997).

\bibitem{Mfox2006}
M. Fox, Quantum optics . (Oxford university Press Inc. New York, 2006).

\bibitem{Emer1970}
E. Merzbacher, Quantum mechanics . (John Wiley and sons, New York, 1970).

\bibitem{Ras1991}
R. A. Swainson and G. W. F. Drake J. Phys. A: Math. Gen \textbf{24} (1991) 95.

\bibitem{Je2004}
J. Evers, U. D. Jentschura and C. H. Keitel, Phys. Rev. A \textbf{70}, 062111 (2004). 

\bibitem{Mhaas2007}
M. Haas et al. Phys. Rev. A \textbf{73} (2007) 052501.
\end{thebibliography}
\end{document}